\documentclass[12pt,a4paper]{article}
\usepackage{tikz,amsmath,amssymb,amsthm}
\usepackage[margin=1.0in]{geometry}
\usepackage{cite}
\usepackage{graphicx}
\usepackage[pagebackref=false]{hyperref}
\usepackage[affil-it]{authblk}
\usetikzlibrary{shapes,snakes}
\usetikzlibrary{decorations.markings}
\usetikzlibrary{decorations.pathreplacing}
\numberwithin{equation}{section}

\def\a{\alpha}

\def\e{\epsilon}

\def\r{\rho}

\newcommand{\inn}{\!\cdot\!}
\newcommand{\nn}{\nonumber}

\def\be{\begin{equation}}
\def\ee{\end{equation}}
\def\bea{\begin{eqnarray}}
\def\eea{\end{eqnarray}}

\makeatletter
\renewcommand\section{\@startsection {section}{1}{\z@}%
                                   {-3.5ex \@plus -1ex \@minus -.2ex}
                                   {2.3ex \@plus.2ex}%
                                   {\normalfont\large\bfseries}}
\renewcommand\subsection{\@startsection{subsection}{2}{\z@}%
                                     {-3.25ex\@plus -1ex \@minus -.2ex}%
                                     {1.5ex \@plus .2ex}%
                                     {\normalfont\bfseries}}
\makeatother

\begin{document}
\begin{titlepage}
{\title{{Gravitational Couplings on D-brane Revisited}}}
\vspace{.5cm}
\author{Ahmad Ghodsi \thanks{a-ghodsi@ferdowsi.um.ac.ir}}
\author{Ghadir Jafari \thanks{ghadir.jafari@stu-mail.um.ac.ir}}
\vspace{.5cm}
\affil{ Department of Physics, Ferdowsi University of Mashhad,    
\hspace{5.5cm} P.O.Box 1436, Mashhad, Iran}
\renewcommand\Authands{ and }
\maketitle
\vspace{-12cm}
\begin{flushright}
\end{flushright}
\vspace{10cm}

\begin{abstract}
Gravitational couplings in bulk space-time include those terms which are fixed by scattering amplitude of strings and ambiguous terms that are coming from the field redefinitions. These field redefinitions can be fixed in the bulk by ghost-free condition. In this paper we have revised the effective gravitational couplings on D-branes by including the field redefinitions. We find the  gravitational effective action up to $\alpha'^2$-order.
\end{abstract}
\end{titlepage}

\section{Introduction}\label{int}
At low energies, string theory exhibits itself  as an effective field theory in space-time. In this regard it can be seen as a sequence of higher derivative corrections that must be added to the lowest energy effective actions for the massless states of closed or open strings. These effective field theories can be obtained or constrained  by a variety of methods including the sigma model, scattering amplitudes or duality considerations \cite{Giveon:1994fu}. These effective actions capture many perturbative properties of the string theory at low energies. 

The same happens in the case of D-branes which are objects in space-time where the open strings end. At low energies, they can be viewed as sub-manifolds where the quantum field theory corresponding to the open strings lives. At the lowest order in the string length, the higher derivative terms can be ignored and as a result, D-branes are completely described by the Dirac-Born-Infeld (DBI) action plus the Wess-Zumino terms \cite{Polchinski:1996na}. 
 
The gravitational terms exist both in the bulk space-time  as well as  on the D-brane world-volume. 
 The leading $\alpha'$-order of the bulk effective action in the Bosonic string theory  is just the known Hilbert-Einstein action
\bea
S^{(0)} =\frac{1}{2\kappa^2}\int d^{D}x \sqrt{-G}R\label{bulk0}\,,
\eea
which is  also shared with the Heterotic and Superstring theories.
The next-to-leading $\alpha'$-order terms of the bulk effective action  have been found in \cite{Metsaev:1987zx} from the corresponding sphere-level S-matrix  elements
\bea
&& S^{(1)} =\frac{\lambda_0}{2\kappa^2}\int d^{D}x e^{\gamma\phi}\sqrt{-G}\ \mathcal{L}_{GB}\label{bulk1}\,,\nn \cr \\
&& \mathcal{L}_{GB}=\alpha'(4 R_{\alpha \beta} R^{\alpha \beta} -  R^2 -  R_{\alpha \beta \mu \nu} R^{\alpha \beta \mu \nu}) \label{GB}\,,
\eea
where $\mathcal{L}_{GB}$ is the Gauss-Bonnet  combination of the curvature squared terms. In the case of Bosonic string theory one has $\lambda_0=-\frac{1}{4}$ while $\lambda_0=-\frac{1}{8}$  and $\lambda_0=0$ are appropriate choices for the Heterotic theory  and  Superstring theory respectively.
However it must be pointed out that unlike the leading $\alpha'$-order action (\ref{bulk0}), the couplings in  (\ref{bulk1}) are not unique and can not  fix by S-matrix considerations completely.  Only the coefficient of the square of the Riemann tensor
is fixed by the string S-matrix, while the other coefficients are arbitrary and they have been fixed to
the above values by a local field redefinition. The Gauss-Bonnet combination is  convenient because up to the quadratic order in 
metric perturbations it is a total derivative and therefore does not change the propagator of gravitons derived from the Hilbert-Einstein action (\ref{bulk0}). As another result of being  total derivative, such combination is also ghost free \cite{ Zwiebach:1985uq }.  See \cite{Tseytlin:1986zz, Metsaev:1986yb } for more  details about the field redefinition ambiguities in the string effective actions.

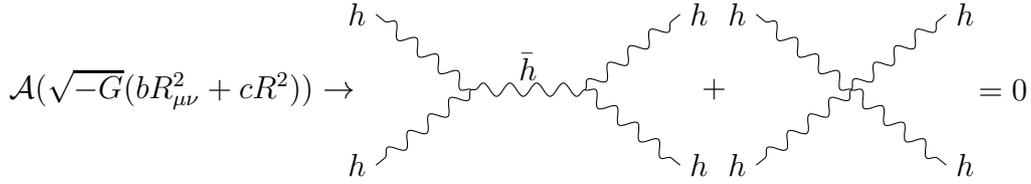
\begin{figure}
	\centering
	\begin{tikzpicture}
		\node at (-2.8,2) {$ {\cal A}(\sqrt{-G}(b R_{\mu\nu}^2+c R^2)) \rightarrow $};
	\draw [decoration={snake},decorate] 
	(2.5,2) -- (3.75,3);
	\draw [decoration={snake},decorate] 
	(2.5,2) -- (3.75,1);
	\draw [decoration={snake},decorate] 
	(2.5,2) -- (1,2);
	\draw [decoration={snake},decorate] 
	(1,2) -- (-.25,3);
	\draw [decoration={snake},decorate] 
	(1,2) -- (-.25,1);
		\node at (1.75,2.3) {$ \bar{h} $};
	\node at (4,3) {$ h $};
	\node at (4,1) {$ h $};
	\node at (-.5,3) {$ h $};
	\node at (-.5,1) {$ h $};
	\node at (4.2,2) {$ + $};
	\draw [decoration={snake},decorate] 
	(6,2) -- (7.25,3);
	\draw [decoration={snake},decorate] 
	(6,2) -- (7.25,1);
	\draw [decoration={snake},decorate] 
	(6,2) -- (4.75,3);
	\draw [decoration={snake},decorate] 
	(6,2) -- (4.75,1);
	\node at (7.5,3) {$ h $};
	\node at (7.5,1) {$ h $};
	\node at (4.5,3) {$ h $};
	\node at (4.5,1) {$ h $};
		\node at (8,2) {$ =0 $};
	\end{tikzpicture}
	\caption{ Exchange diagram and contact diagram cancellation. In above diagram $h$ denotes the fluctuation fields coming from $R_{\mu\nu}^2$ or $R^2$ terms and $\bar{h}$ denotes the usual graviton propagator.}\label{bulk2g}
\end{figure}

From the field theory point of view the ambiguities in the coefficients of $R_{\mu\nu}^2$ and $R^2$ come from the exact cancellation between the exchange diagram and contact diagram contribution of the field theory amplitude \cite{Deser:1986xr}. In fact by considering a general Lagrangian at $ \alpha' $-order in the following form
\begin{equation}
{\cal L}(\alpha')=\alpha'\sqrt{-G}[a R_{\mu\nu\alpha\beta}R^{\mu\nu\alpha\beta}+b R_{\mu\nu}R^{\mu\nu}+c R^2 ]\,,
\end{equation}
and by comparing it with the low energy string amplitude results, one can only fix the ``$a$'' coefficient in the above Lagrangian 
and the other contributions from $ R_{\mu\nu}^2 $ and $ R^2 $ terms  cancel each other exactly, independent of the values for $b$ or $c$ (see Fig.\ref{bulk2g}).

It is interesting that such cancellation also happens beyond the tree level calculations so that the higher genus string calculations also can not fix these ambiguous terms \cite{Forger:1996vj}. It is stated that this property means that the field redefinitions are a symmetry of the perturbative string S-matrix \cite{Tseytlin:1986zz, Metsaev:1986yb }. As a result of this symmetry, those terms where their coefficients can be changed by some local field  redefinitions are ambiguous. In the above case one can show that by a 
 general field redefinition of the metric at $ \alpha' $-order as
\begin{equation}
G'_{\mu\nu}=G_{\mu\nu}+\alpha'(d_1 R_{\mu\nu}+d_2 G_{\mu\nu}R)\,,  \label{Frdef}
\end{equation}
where we consider $ \alpha' $ to be small, and applying this to the Hilbert-Einstein term
\eqref{bulk0}, up to $ \alpha' $-order one can produce the following terms
\begin{equation}
{\cal L}'(\alpha')=\alpha'\sqrt{-G}[-d_1 R_{\mu\nu}R^{\mu\nu}+(\tfrac{d_1}{2}+\tfrac{1}{2} d_2(D-2)) R^2 ]\,.
\end{equation}
Consequently by an appropriate choice of $ d_1=1$ and $ d_2=\frac{-1}{2D-4} $ one can get the Gauss-Bonnet  combination which has the advantage of being a ghost-free Lagrangian.  

At $\alpha'^2$-order we must have an action with six number of derivatives as studied in  \cite{ Metsaev:1986yb}. Similarly at this order, terms containing the Ricci tensor are ambiguous. There are just two unambiguous independent $R^3$ structures
\bea
I_1=R_{\alpha \beta}{}^{\mu \nu} R^{\alpha \beta \kappa \lambda} R_{\kappa \lambda \mu \nu}
\,, \ I_2=R_{\alpha}{}^{\mu}{}_{\kappa}{}^{\nu} R^{\alpha \beta \kappa \lambda} \
R_{\beta \mu \lambda \nu} \,,
\eea
where the coefficients has been fixed as follows\cite{ Metsaev:1986yb}
\bea
S^{(2)} =\frac{{\alpha'}^2}{48\kappa^2}\int d^{D}x e^{2\gamma\phi}\sqrt{-G}\ (3I_1-4I_2)\,. \label{Bulk2}
\eea
Applying the field redefinition  \eqref{Frdef} with the mentioned fixed coefficients to the following action where we have fixed it just by S-matrix calculations 
\begin{equation}\label{easmatrix1}
{\cal L}(\alpha')=\sqrt{-G}[R+\frac{\alpha'}{4} R_{\mu\nu\alpha\beta}R^{\mu\nu\alpha\beta}]\,,
\end{equation}
we can produce the following $ \alpha'^2 $-order terms
\bea
\mathcal{L}'({\alpha'}^2)={\alpha'}^2\sqrt{-G} (I_3+ \frac{3 (D-1)}{16 (D-2)} R \square R -  \tfrac{3}{4} R^{\alpha \beta} \square R_{\alpha \beta})\,,\label{BB}
\eea
where
\bea
I_3&=&\tfrac{3}{2} R_{\alpha}{}^{\gamma} R^{\alpha \beta} R_{\beta \gamma} - \frac{D-1 }{2 (D-2)}R_{\alpha \beta} R^{\alpha \beta} R + \frac{D+4 }{32 (D-2)}R^3 -  \tfrac{1}{2} R^{\alpha \beta} R^{\gamma \delta} R_{\alpha \gamma \beta \delta} \nn \cr \\ 
&+& \frac{D}{16 (D-2)} R R_{\alpha \beta \gamma \delta} R^{\alpha \beta \gamma \delta}-  \tfrac{1}{2} R^{\alpha \beta} R_{\alpha}{}^{\gamma \delta \eta} R_{\beta \gamma \delta \eta}\,.
\eea
We can check that the last two terms in equation (\ref{BB}) lead to the ghost modes at this order. To prevent this and to have a ghost-free action we must supplement the field redefinition \eqref{Frdef} with the following extra terms i.e.
\footnote{In fact for the action to be ghost-free at all order of $ \alpha' $ one can show that
the general field redefinition is \cite{Jack:1987pc,Lawrence:1988rb}, 
 $G'_{\mu\nu}=G_{\mu\nu}+\sum_{n=1}^{\infty} \alpha'^n( a_n \square^{n-1}R_{\mu\nu}+b_n G_{\mu\nu}\square^{n-1} R).$}
\begin{equation}
G'_{\mu\nu}=G_{\mu\nu}+\alpha'(R_{\mu\nu}-\frac{1}{2D-4} G_{\mu\nu}R) +\alpha '^2( - \tfrac{3}{4}  \square R_{\mu\nu} + \frac{3 (D-3) }{8 (D-2)^2}G_{\mu\nu} \square R)\label{Frdef1}\,.
\end{equation}
Consequently the full action at this order will be
\bea
S^{(2)} =\frac{{\alpha'}^2}{48\kappa^2}\int d^{D}x e^{2\gamma\phi}\sqrt{-G}\ (3I_1-4I_2+ 48 I_3)\label{bulk2}\,.
\eea
In next sections we will apply the same technique to find the effective gravitational action on D-branes. We will explain that how the field redefinitions and actions  for the bulk space-time will be useful to construct  actions on D-branes.

The outline for the remaining parts of the paper is arranged as follow: In section 2 we review the effective action on D-branes at $\a'$-order and we find new extra  terms that must be added due to the field redefinitions. In section 3 we do the same steps as in section 2 but for $\a'^2$-order of the effective action. To do this we need the tree-level string amplitude computation corresponding to scattering of two massless closed strings from a D-brane. Using the results of previous sections we will find an action which correctly reproduces the string theory amplitude at this order. In last section  we will summarize and discuss our results.

\section {On \texorpdfstring{$\a'$}{Lg} corrections to \texorpdfstring{$D_p$}{Lg}-brane action}
There is a similar story as bulk space-time for the gravitational corrections to the $D_p$-brane world-volume. The leading $\alpha'$-order (DBI action) for $D_p$-brane is \cite{Leigh:1989jq,Bachas:1995kx}
\bea
S^{(0)}_{D_p}=-T_p\int d^{p+1}x
e^{-\phi}\sqrt{-det(\tilde{G}_{ab})}\,,\label{dbi}
\eea
where $\tilde{G}_{ab}$  is the pull-back of the bulk metric $G_{\mu\nu}$ onto the world-volume of the $D_p$-brane
\bea
\tilde{G}_{ab}=\frac{\partial X^{\mu}}{\partial{\sigma^a}}\frac{\partial X^{\nu}}{\partial{\sigma^b}}G_{\mu\nu}\,.
\eea
The $\alpha'$-order  gravitational corrections to this action in Bosonic string theory has been constructed by using the consistency of the effective action with  $\alpha^{\prime}$-order
terms of the  disk-level scattering amplitude of two gravitons from $D_p$-branes \cite{Corley:2001hg}. These corrections are\footnote{Our index notation is such that  the Greek letters are used for space-time indices,
	the Latin letters $(a,b,c,...)$ for the world-volume indices and $(i,j,k,...)$ for 
	the transverse or normal bundle indices.}
\bea
S_{D_p}^{(1)}=
-\frac{ \alpha'T_p}{2}\int d^{p+1}xe^{-\phi}
\sqrt{-\tilde{G} }\bigg[\tilde{R}+(\Omega^i{}_{ a}{}^a
\Omega_{i\, b}{}^b-\Omega^{i}_{ab}
\Omega_{i}^{ab}) \bigg]\label{dbi1}\,,
\end{eqnarray}
where $\tilde{R}$ is the scalar curvature constructed from the pull-back  metric  $\tilde{G}_{ab}$ and 
 $\Omega$ is the second fundamental form.
It should be noted that using the Gauss-Codazzi identity\cite{Carter:1997pb,Bachas:1999um},   
$\tilde{R}_{abcd}=R_{abcd}+\delta_{ij}(\Omega^i{}_{ac}\Omega^j{}_{db}-{\Omega^{i}}_{ad}{\Omega^{j}}_{cb})$, we can rewrite the  Lagrangian in the above action as
  \begin{equation} \label{Lagbos}
    \mathcal{L}_{D_p}({\alpha'}) =2\tilde{R}- 	R_{abcd}\tilde{G}^{ac}\tilde{G}^{bd}=2\tilde{R}- R_{\alpha\mu\beta\nu}\tilde{G}^{\alpha\beta}\tilde{G}^{\mu\nu}\,,
    \end{equation}
where $ R_{abcd} $ is the pullback of the Riemann tensor in the Bulk
and  $\tilde{G}^{ab}$ is the inverse of the Pullback metric. The second equality in Eq. (\ref{Lagbos}) is coming from replacing the pullback Riemann tensor with the bulk one. In this form, $\tilde{G}^{\alpha\beta}$  is the first fundamental tensor i.e. $
\tilde{G}^{\mu\nu}=\frac{\partial X^{\mu}}{\partial{\sigma^a}}\frac{\partial X^{\nu}}{\partial{\sigma^b}}\tilde{G}^{ab}$.

However it must be noted that, in writing the $D_p$-brane action there are ambiguities in terms which are constructed out of the  Ricci tensor. We can see this by varying the  DBI action 
$\delta(\sqrt{-\tilde{G}})=\frac12 \sqrt{\tilde{-G}}\,\tilde{G}^{ab}\delta\tilde{G}_{ab}$ 
and choosing $ \delta\tilde{G}_{ab}=\alpha'(d_1 R_{ab}+d_2 \tilde{G}_{ab}R)  $ as a field redefinition. Then one gets the following terms
\begin{equation}\label{Frdefbrane}
\delta{\cal L}({\alpha'})=\frac{\alpha'}{2} [d_1 R_{\alpha\beta}\tilde{G}^{\alpha\beta}+d_2(p+1) R]\,.
\end{equation}
 So in general we must add these terms to the action \eqref{dbi1}. In \cite{Corley:2001hg} the coefficients of these terms have been chosen to zero.

It's worth it to note that the field redefinition here can be constructed from the field redefinition in the bulk (\ref{Frdef}) by a simple pullback. Therefore to have a unique field redefinition for both bulk and $D_p$-brane space-time we must choose the values of $d_1=1$ and $d_2=\frac{-1}{2D-4}$. Finally the $\alpha'$-order effective Lagrangian  on $D_p$-branes must be
\be\label{EXlag1}
\mathcal{L}_{D_p}({\alpha'})=2\tilde{R}- R_{\alpha\mu\beta\nu}\tilde{G}^{\alpha\beta}\tilde{G}^{\mu\nu}+ R_{\alpha\beta}\tilde{G}^{\alpha\beta}-\frac{p+1}{8(D-2)} R\,.
\ee

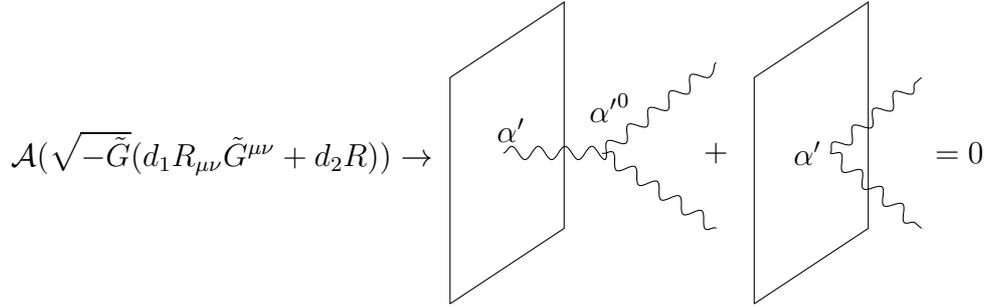
\begin{figure}
 	\centering
 	\begin{tikzpicture}
 		\node at (-1.5,2) {$ {\cal A}(\sqrt{-\tilde{G}}(d_1 R_{\mu\nu}\tilde{G}^{\mu\nu}+d_2 R))\rightarrow $};
 	\draw (1.5,0) -- (3,1) -- (3,4) -- (1.5,3) -- (1.5,0);
 	\draw [decoration={snake},decorate] 
 	(2.2,2) -- (3.5,2) -- (5,3.2);
 	\draw [decoration={snake},decorate]  (3.5,2) -- (5,1);
 	\node at (2.3,2.3) {$\alpha'$};
		\node at (3.6,2.6) {${\alpha'}^0$};
 		\node at (5,2) {$ + $};
 	\draw (5.5,0) -- (7,1) -- (7,4) -- (5.5,3) -- (5.5,0);
 	\draw [decoration={snake},decorate] 
 	(6.5,2) -- (7.7,3);
 	\draw [decoration={snake},decorate] 
 	(6.5,2) -- (7.7,1);
 	\node at (6.2,2) {$\alpha'$};
 	\node at (8.2,2) {$ =0 $};
 	\end{tikzpicture}
 	\caption{\small{$t$-channel and contact term cancellation  for ambiguous terms in  the D-brane action.}}\label{DbraneAmbg}
 \end{figure}

Moreover, because the field redefinition here is coming from the field redefinition of the bulk, we may expect that a similar situation happens for cancellation between exchange and contact diagrams. In $D_{p}$-brane case, the last two terms in Eq. (\ref{EXlag1}) produce both t-channel (by imposing a source at $\alpha'$-order) and contact term contributions (see Fig. \ref{DbraneAmbg}) which cancel each other exactly. In computing the t-channel the vertex of the bulk is coming from bulk action (\ref{bulk0}) at order ${\alpha'}^0$ i.e.
\begin{eqnarray}
\!\!\!\!\!\!\!\!\!\!\!\!&&(\mathcal{V}^{\alpha'^{0}}_{h\e_1\e_2})^{\alpha\beta}=-3 (k_1.{\epsilon_2}.k_1)  \epsilon_1^{\alpha \beta} + \frac32 t (\epsilon_1.\epsilon_2)^{\alpha \beta}  - \frac32 (k_2.\epsilon_1.\epsilon_2.k_1) \eta^{\alpha \beta}-  \tfrac{9}{8} t Tr(\epsilon_1.\epsilon_2)  \eta^{\alpha \beta} \nonumber \\
\!\!\!\!\!\!\!\!\!\!\!\!&& - 3 Tr(\epsilon_1.\epsilon_2) k_1^{\alpha} k_1^{\beta} + 6 k_1^{\alpha} (k_1.\epsilon_2.\epsilon_1^{\beta}) \!-\! \frac32 Tr(\epsilon_1.\epsilon_2) k_1^{\alpha} k_2^{\beta}  + 3  (k_2.\epsilon_1^{\alpha})(k_1.\epsilon_2^{\beta})+(1\leftrightarrow 2) \,.\label{V0}
\end{eqnarray}
As an example if we consider $ \sqrt{-\tilde{G}} R_{\alpha\beta}\tilde{G}^{\alpha\beta} $, it produces a source term of $\a'$-order as
\begin{equation}
(\mathcal{S}^{\alpha'}_{h})^{\alpha\beta}=-\tfrac{1}{2}k^2\, V^{\alpha\beta}-\tfrac{1}{2}(k.V.k) \eta^{\alpha\beta}+k^{\alpha} (k.V^{\beta})\label{S1}\,.
\end{equation}
Now by using Eq. \eqref{V0} and Eq. \eqref{S1} and the graviton propagator
\begin{equation}
  (\mathcal{P}_{hh})_{\mu\nu\alpha\beta}=- \dfrac{i}{2 k^2} ({\eta_{\alpha\nu} \eta_{\beta\mu} + \eta_{\alpha\mu} \eta_{\beta \nu} -  \frac{2}{D-2} \eta_{\alpha \beta} \eta_{\nu \mu}})\,,
  \end{equation}
 one can compute the amplitude below
\begin{eqnarray}
&&{\cal A}_t=\frac{-t}{t}( k_1 \cdotp V \cdotp {\epsilon_2} \cdotp {\epsilon_1} \cdotp k_2 
-k_1 \cdotp V \cdotp {\epsilon_1} \cdotp {\epsilon_2} \cdotp k_1    
+ \tfrac{1}{2}  k_1\cdotp {\epsilon_2}\cdotp k_1 Tr( \epsilon_1 \cdotp V) - k_1\cdotp {\epsilon_2}\cdotp V\cdotp      {\epsilon_1}\cdotp k_2  \nonumber \\ 
&& + \tfrac{1}{2} k_2 \cdotp {\epsilon_1}\cdot k_2 Tr(\epsilon_2\cdotp V) 
-  \tfrac{1}{2} s\, Tr(\epsilon_1 \cdotp {\epsilon_2})  -  \tfrac{1}{2} t\, Tr(\epsilon_1 \cdotp V \cdotp {\epsilon_2})\,, \label{ATT1}
\end{eqnarray}
where  $t$ and $s$ are the Mandelstam  variables defined such that $t=-2k_1 \cdotp k_2$  and $s=-{\frac12}k_1 \cdotp D \cdotp k_1$ (for more details see the next section).
By expansion of $ \sqrt{-\tilde{G}} R_{\alpha\beta}\tilde{G}^{\alpha\beta} $ to second order in perturbations and going to the momentum space we find the contact terms contribution
\begin{eqnarray}
&&{\cal A}_c=k_1 \cdotp V \cdotp {\epsilon_2} \cdotp {\epsilon_1} \cdotp k_2 
-k_1 \cdotp V \cdotp {\epsilon_1} \cdotp {\epsilon_2} \cdotp k_1    
+ \tfrac{1}{2}  k_1\cdotp {\epsilon_2}\cdotp k_1 Tr( \epsilon_1 \cdotp V) - k_1\cdotp {\epsilon_2}\cdotp V \cdotp      {\epsilon_1}\cdotp k_2  \nonumber \\ 
&& + \tfrac{1}{2} k_2 \cdotp {\epsilon_1}\cdot k_2 Tr(\epsilon_2\cdotp V) 
-  \tfrac{1}{2} s\, Tr(\epsilon_1 \cdotp {\epsilon_2})  -  \tfrac{1}{2} t\, Tr(\epsilon_1 \cdotp V \cdotp {\epsilon_2})\,,
\end{eqnarray}
which exactly cancel the contribution of \eqref{ATT1}. The same scenario exists for $\sqrt{\tilde{G}} R$.

Note that it can not be possible to have s-channel at $\alpha'$-order for these last two terms in Eq. (\ref{EXlag1}) because they  do not produce open-closed vertex at  this order.
We will show in next section that these terms are necessary for consistency of the bulk and $D_p$-brane actions with the string theory S-matrix at $ \a'^2 $-order. This is because any choose of field redefinition must be used for both bulk and $D_p$-brane identically. 
 
\section{On \texorpdfstring{$ {\alpha'}^2 $}{Lg} corrections to \texorpdfstring{$D_p$}{Lg}-brane action}
In this section we are going to go one step further and find the  $ {\alpha'}^2 $-order of effective gravitational action on $D_p$-branes with using the bulk and D-brane actions we found in previous section. We must take into account the field redefinitions  too. But before we go through the action we need to compute the amplitude coming from the S-matrix calculations. 
\subsection{The string theory amplitude}\label{stamp}
In Bosonic string theory the effective gravitational actions on $D_{p}$-branes have been computed from S-matrix  for the first and next leading order in \cite{Corley:2001hg}, so we ignored the details of this computation in previous section. But to compute the $ {\alpha'}^2 $-order we need to expand the scattering amplitude one more order, therefore we present the details of our calculations in this section.

The disk-level Bosonic string scattering amplitude of two closed strings massless states off a $D_{p}$-brane is calculated in \cite{Corley:2001hg} and is given by the following expression 
\bea
{\cal A}_{D_2}(k_1, \e_1 \!\!\!\! &;  & \!\!\!\! k_2, \e_2) =\frac{i}{4} \, g^{2}_c
\, C_{D_2}
\, (2 \pi)^{p+1} \, \delta^{p+1}(k_1 + k_2) \, \Bigl( d_1 \, B(-\frac{t}{2}, 
2s+1) + d_2 \, B(-\frac{t}{2},2 s) \nonumber \\
& & - \, d_3 \, B(1-\frac{t}{2}, 2s) + d_4 \, B(1-\frac{t}{2},2s+1)
+ d_5 \, B(-1-\frac{t}{2},2s+1) \nonumber \\
& &  + \, d_6 \, B(1-\frac{t}{2}, 2s-1) + d_7 \, B(-1-\frac{t}{2},2s-1) 
- d_8 \, B(-\frac{t}{2},2s-1) \nonumber \\
& &  - \, d_9 \, B(2-\frac{t}{2},2s-1) + d_{10} \, B(3-\frac{t}{2},2s-1)
\Bigr) \,,
\label{exactamp}
\eea
where all coefficients for graviton polarization are given in the appendix A. \footnote{In the above amplitude $\alpha'$ has been set to 2 for convenience, it can be recovered everywhere by dimensional analysis.} 
The Mandelstam  variables $t$ and $s$ are defined such that $t=-2k_1 \cdotp k_2$ is the square  of the momentum transfer in the transverse directions and $s=-{\frac12}k_1 \cdotp D \cdotp k_1$ is the momentum flow parallel to the world-volume of the $D_p$-brane.
In the above expressions $D^{\mu \nu} =2 V^{\mu \nu} - \eta^{\mu \nu}$
where $V^{\mu \nu} = \mbox{diag}(-1,1,...,1,0,...,0)$ is the flat metric along the $D_p$-brane directions, and $\eta^{\mu \nu}$ is just the $D$-Dimensional Minkowski metric.
In this notation  all indices are contracted with the Minkowski metric $\eta_{\mu \nu}$.
We refer the interested readers to \cite{Corley:2001hg} for further reading.

 In order to find the effective actions,  we need to expand Eq. (\ref{exactamp}) in powers  of $\a' k^2$ . The $(\alpha' k^2)^0$ expansion precisely gives rise to the DBI action (\ref{dbi}) and the $\alpha' k^2$-order is produced by (\ref{dbi1})\cite{Corley:2001hg}. 
Here we expand the above amplitude and keep the terms at order $ (\alpha'k^2)^2  $. Doing this, we can observe that there are three types of contribution to the whole amplitude: closed and open string poles (t ans s channels) as well as contact interactions (contact terms)
\bea
{\cal A}(\alpha'^2)={\cal A}_{t}(\alpha'^2)+{\cal A}_{s}(\alpha'^2)+{\cal A}_{c}(\alpha'^2).
\eea
The closed string pole  is given just by  one term
\bea
{\cal A}_{t}(\alpha'^2)=-\frac{1}{8t} s (k_1 \cdotp \e_2 \cdotp k_1)(k_2 \cdotp \e_1 \cdotp k_2)\,. \label{ATo2}
\eea  
Similarly the $s$-channel amplitude will be
\bea \label{ASo2}
{\cal A}_{s}(\alpha'^2)=-\frac{1}{8s} t (k_1 \cdotp V \cdotp \e_2 \cdotp V \cdotp k_1)(k_1 \cdotp V \cdotp \e_1 \cdotp V \cdotp k_1)\,.
\eea
The remaining terms construct the contact interactions
\bea
&&{\cal A}_{c}(\alpha'^2)= \tfrac{1}{192} s \bigl(2 (6 + \pi^2) s + 3 t\bigr) \mbox{Tr}(\epsilon_1 \cdotp \epsilon_2) 
- \tfrac{1}{48} \bigl(2 (6 + \pi^2) s + 3 t\bigr) k_1 \cdotp V \cdotp \epsilon_2 \cdotp \epsilon_1 \cdotp k_2  \nn \\ 
&&+ \tfrac{1}{192} \bigl(2 (\pi^2-12) s t-48 s^2 - 3 t^2\bigr) \mbox{Tr}(\epsilon_1 \cdotp V \cdotp \epsilon_2) -  \tfrac{1}{8} (k_1 \cdotp \epsilon_2 \cdotp V \cdotp k_1)(k_2 \cdotp \epsilon_1 \cdotp k_2) 
\nonumber \\ 
&&+ \tfrac{1}{192} \bigl(48 s^2 - 2 (\pi^2-12) s t  + 3 t^2\bigr) \mbox{Tr}(\epsilon_1 \cdotp V \cdotp \epsilon_2 \cdotp V)  
+ \tfrac{1}{8} (k_1 \cdotp V \cdotp \epsilon_2 \cdotp V \cdotp k_1)(k_2 \cdotp \epsilon_1 \cdotp k_2)\nonumber \\ 
&&- \tfrac{1}{48} \bigl((\pi^2-3) t-12 s\bigr)k_2 \cdotp V \cdotp \epsilon_1 \cdotp \epsilon_2 \cdotp V \cdotp k_1 
+ \tfrac{1}{96} \bigl(2 (\pi^2-6) s - 3 t\bigr) k_2 \cdotp \epsilon_1 \cdotp V \cdotp \epsilon_2 \cdotp k_1 \nonumber \\ 
&& -  \tfrac{1}{48} \pi^2 s  \mbox{Tr}(\epsilon_1 \cdotp V)(k_1 \cdotp \epsilon_2 \cdotp k_1) 
 + \tfrac{1}{384} \bigl( 4(\pi^2-6) s t-96 s^2 + \pi^2 t^2\bigr) \mbox{Tr}( \epsilon_1 \cdotp V)\mbox{Tr}(\epsilon_2 \cdotp V) \nonumber \\ 
&&-  \tfrac{1}{48} \pi^2 t \mbox{Tr}(\epsilon_1 \cdotp V)(k_1 \cdotp \epsilon_2 \cdotp V \cdotp k_1 ) 
+ \tfrac{1}{48} \bigl((\pi^2-6) t-24 s\bigr) \mbox{Tr}(\epsilon_1 \cdotp V) (k_1 \cdotp V \cdotp \epsilon_2 \cdotp V \cdotp k_1) \nonumber \\ 
&& + \tfrac{1}{8} (4 s + t) k_1 \cdotp V \cdotp \epsilon_2 \cdotp V \cdotp \epsilon_1 \cdotp k_2 -\tfrac12 \bigl(s - \tfrac{1}{24} ( \pi^2-6) t\bigr) k_2 \cdotp V \cdotp \epsilon_1 \cdotp V \cdotp \epsilon_2 \cdotp V \cdotp
k_1 \nonumber \\ 
&&+\tfrac{1}{4} (k_2 \cdotp \epsilon_1 \cdotp V \cdotp k_1)(k_1 \cdotp \epsilon_2 \cdotp V \cdotp k_2)  
+ \tfrac{1}{2} (k_1 \cdotp V \cdotp \epsilon_1 \cdotp k_2)(k_1 \cdotp V \cdotp \epsilon_2 \cdotp V \cdotp k_1) +(1\leftrightarrow 2)\,.\label{AC1}
\eea
There are some interesting points to compare this amplitude with similar amplitude in Superstring theory. The first point is that  the $\alpha'^2$ part of the amplitude for scattering of a Superstrings from $D_p$-brane as computed in  \cite{Garousi:1996ad} just contains contact interactions whereas here we have closed and open string poles as well. Another interesting point is that if we take a look at the above contact terms, we see that some terms contain $\pi^2$ as coefficient and others do not. If we separate these terms, we recover the $\alpha'^2$-order of the Superstring amplitude.
The Superstring amplitude is  \cite{Garousi:1996ad}
\begin{equation}
     \mathcal{A}=-\frac{ T_p}{2}\,
     \frac{\Gamma(-t/2)\Gamma(-2s)}{\Gamma(1-t/2-2s)}
     \left(-2s\,a_1+\frac{t}{2}\,a_2\right) ,
     \label{finone}
     \end{equation}
     where the kinematic factors $a_1$ and $a_2$  are given by
     \begin{eqnarray}
         &&a_1={ Tr}(\epsilon_1\inn D)\,k_1\inn \epsilon_2 \inn k_1 -k_1\inn\epsilon_2\inn
     D\inn\epsilon_1\inn k_2 -2 k_1\inn\epsilon_2\inn\epsilon_1 \inn D\inn k_1
     \nonumber \\
     &&- k_1\inn\epsilon_2\inn \epsilon_1\inn k_2-
     s\,{\rm Tr}(\epsilon_1\inn\epsilon_2)
     +(1\leftrightarrow 2)\,,\cr
        \nonumber \\
     &&a_2=s\,{\rm Tr}(\epsilon_1\inn\epsilon_2)+{\rm Tr}(\epsilon_1\inn D)\,(2k_1\inn\epsilon_2\inn D\inn k_2 +k_2\inn D\inn\epsilon_2\inn D\inn k_2)
     +k_1\inn D\inn\epsilon_1\inn D\inn\epsilon_2\inn D\inn k_2
     \nonumber \\
     &&-k_1\inn D\inn\epsilon_1\inn\epsilon_2\inn
     D\inn k_2
     -s\,{\rm Tr}(\epsilon_1\inn D\inn \epsilon_2\inn D)
          +{\rm Tr}(\epsilon_1\inn D) {\rm Tr}(\epsilon_2\inn D)\,(s+\frac{t}{4})
     +(1\leftrightarrow 2)\,. 
     \label{finthree}\nonumber
\end{eqnarray}
 The $\alpha'^2$-order of this amplitude is
\bea
\mathcal{A}({\a'}^2)=-\frac{\pi^2\alpha'^2T_p}{48}  (-2s\,a_1+\frac{t}{2}\,a_2)\,. \label{a2Amp}
\eea
These terms are exactly the same terms in (\ref{AC1}) that have $\pi^2$ in their coefficients. This feature has been studied  recently in \cite{Huang:2016tag} that $\alpha'$ expansion of Bosonic string amplitudes has the same transcendental pieces as found for the Superstring. Our calculations also support their conjecture of {\it{uniform transcendentality}} in string theories. In the next section we will show that the Bosonic effective action at this order  is the Superstring one in (\ref{SupD2})  plus some additional terms with rational coefficients instead of $\pi^2$.

\subsection{Field theory calculations}\label{fiamp} 

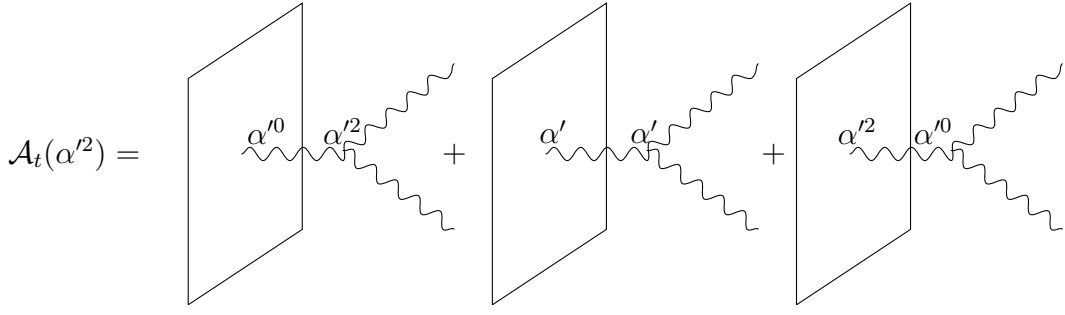
\begin{figure}
  	\centering
  	\begin{tikzpicture}
  	\node at (0,2) {$ {\cal A}_{t}(\alpha'^2) =$};
  	\draw (1.5,0) -- (3,1) -- (3,4) -- (1.5,3) -- (1.5,0);
  	\draw [decoration={snake},decorate] 
  	(2.2,2) -- (3.5,2) -- (5,3.2);
  	\draw [decoration={snake},decorate]  (3.53,2.05) -- (5,1);
  	\node at (2.53,2.3) {$\alpha'^0$};
		\node at (3.53,2.3) {$\alpha'^2$};
  	\node at (5,2) {$ + $};
  	\draw (5.5,0) -- (7,1) -- (7,4) -- (5.5,3) -- (5.5,0);
  	\draw [decoration={snake},decorate] 
  	(6.2,2) -- (7.5,2) -- (9,3.2);
  	\draw [decoration={snake},decorate]  (7.53,2.05) -- (9,1);
  	\node at (6.3,2.3) {$\alpha'$};
  	\node at (7.5,2.3) {$\alpha'$};
  	\node at (9.2,2) {$ + $};
  	\draw (9.5,0) -- (11,1) -- (11,4) -- (9.5,3) -- (9.5,0);
  	\draw [decoration={snake},decorate] 
  	(10.2,2) -- (11.5,2) -- (13,3.2);
  	\draw [decoration={snake},decorate]  (11.53,2.05) -- (13,1);
  	\node at (10.3,2.3) {$\alpha'^2$};
			\node at (11.3,2.3) {$\alpha'^0$};
  	\end{tikzpicture}
  	\caption{\small{Field theory $t$-channel diagrams at $ \alpha'^2 $}}\label{talpha2}
  \end{figure}

Given the string scattering amplitude presented in the previous section, we can find the low energy effective action for the $D_{p}$-brane that reproduces them. 
The Feynman diagrams for the   amplitude in t-channel is depicted in  (Fig.\ref{talpha2}). 
The field theory amplitude in t-channel  can be computed by
  \begin{equation}
  \mathcal{A}_t(\e_1\e_2)=(S_{h})^{\mu\nu} (\mathcal{P}_{hh})_{\mu\nu\alpha\beta} (\mathcal{V}_{h\e_1\e_2})^{\alpha\beta}, \label{ATf}
  \end{equation} 
  where $ (\mathcal{P}_{hh})_{\mu\nu\alpha\beta} $  is the propagator of gravitons, $  (S_{h})^{\mu\nu} $ is the graviton source on the $D_p$-brane that comes from the linear expansion of the effective action on $D_p$-brane and $ (\mathcal{V}_{h\e_1\e_2})^{\alpha\beta} $ is the three graviton vertex on the bulk.
	
There are three possible low energy diagrams in this channel at ${\a'}^2$-order as depicted in (Fig.\ref{talpha2}): 

$\bullet$ In first diagram the source is coming from the DBI action on $D_p$-brane
\begin{equation}
(\mathcal{S}^{\alpha'^{0}}_{h})^{\alpha\beta}=\tfrac{1}{2}\tilde{G}^{\alpha\beta}\,,
\end{equation}
and the vertex in the bulk comes from the action \eqref{bulk2}
\begin{eqnarray}
&&(\mathcal{V}^{\alpha'^2}_{h\e_1\e_2})^{\alpha\beta}=\tfrac{1}{8} t^3 (\epsilon_1.\epsilon_2)^{\alpha\beta} 
-\frac{t^3 Tr(\epsilon_1.\epsilon_2) \eta^{\alpha \beta}}{2(D-2)} 
+ \tfrac{1}{2} t^2  k_2^{\alpha} (k_1.\epsilon_2.\epsilon_1)^{\beta}+ \tfrac{1}{4} t^2  (k_2.\epsilon_1)^{\alpha}  (k_1.\epsilon_2)^{\beta}   \nonumber \\ 
&& 
+ \frac{(D-6) t^2 Tr(\epsilon_1.\epsilon_2) k_1^{\alpha} k_1^{\beta}}{4 (D-2)}
-  \frac{2 t^2 (k_1.\epsilon_2.\epsilon_1.k_2) \eta^{\alpha \beta}} {D-2}
-\frac{(D+2) t^2 Tr(\epsilon_1.\epsilon_2) k_1^{\alpha} k_2^{\beta}}{4 (D-2)}  \nonumber \\ 
&& + \frac{(D-6) t (k_2.\epsilon_1.\epsilon_2.k_1) k_1^{\alpha} k_1^{\beta} }{D-2 }
- \frac{(D+6) t (k_2.\epsilon_1.\epsilon_2.k_1) k_1^{\alpha} k_2^{\beta} }{2(D-2)}+ t  ( k_1.\epsilon_2. k_1) k_2^{\alpha} (k_2.\epsilon_1)^{\beta} \nonumber \\ 
&& +\frac{(k_2.\epsilon_1.k_2) (k_1.\epsilon_2.k_1)}{D-2}\big(-  2 t  \eta^{\alpha \beta} 
 + (D-6) k_1^{\alpha} k_1^{\beta}
 -  4  k_1^{\alpha} k_2^{\beta}\big)+(1\leftrightarrow 2) \,.
\end{eqnarray}

 Computing (\ref{ATf})  produces the t-channel amplitude in Eq. \eqref{ATo2} plus some additional contact terms.

$\bullet$ In second diagram the source of gravitons on the $D_p$-brane is at order $ \alpha' $. The action \eqref{dbi1} does not produce any source because at linear order it is total derivative. But those terms in Eq. \eqref{Frdefbrane} that have been produced by field redefinition lead to some sources (for example see Eq. (\ref{S1})). Using the three graviton vertex diagram from the GB terms \eqref{bulk1} in the bulk which has been created by the same field redefinition 
\begin{eqnarray}
&&  (\mathcal{V}^{\alpha'}_{h\e_1\e_2})^{\alpha\beta}={\alpha'} \big(- \tfrac{3}{4} t^2 (\epsilon_1.\epsilon_2)^{\alpha \beta} + \tfrac{3}{2} (k_1.\epsilon_2.k_1) (k_2.\epsilon_1.k_2) \eta^{\alpha \beta} + \tfrac{3}{2}t (k_2.\epsilon_1.\epsilon_2.k_1)  \eta^{\alpha \beta} \nonumber \\
&& + \tfrac{3}{8}t^2 Tr(\epsilon_1.\epsilon_2)  \eta^{\alpha \beta} - 3 (k_2.\epsilon_1.k_2) k_1^{\alpha} k_1.\epsilon_2^{\beta}-  \tfrac{3}{2} t\,  k_2^{\alpha}\, k_1.\epsilon_2.\epsilon_1^{\beta} + 3 (k_2.\epsilon_1.\epsilon_2.k_1) k_1^{\alpha} k_2^{\beta}  \nonumber\\
&&+ \tfrac{3}{2} t\, Tr(\epsilon_1.\epsilon_2) k_1^{\alpha} k_2^{\beta} -  \tfrac{3}{2} t\, k_2.\epsilon_1^{\alpha}\, k_1.\epsilon_2^{\beta}- 3 (k_1.\epsilon_2.k_1) k_2^{\alpha}\, k_2.\epsilon_1^{\beta} -  \tfrac{3}{2} t\, k_1^{\alpha}\, k_2.\epsilon_1.\epsilon_2^{\beta}\big)\,,
\end{eqnarray}
and by computing (\ref{ATf}) on can show that this diagram will not produce any pole and just lead to some contact terms. 

$\bullet$ For the last diagram we need a source for graviton at $ \alpha'^2 $-order. There are two type of terms here. Terms such as 
$ \tilde{G}^{\alpha\beta}\nabla_{\alpha}\nabla_{\beta}R $, $ \square \tilde{R} $ and $  \tilde{\square} R$ become total derivative at linear expansion and do not produce any source. Moreover there are other terms for example $ \square R_{\alpha\beta} \tilde{G}^{\alpha\beta}$  which gives rise to the following source 
\begin{equation}
(\mathcal{S}^{\alpha'^2}_{h})^{\alpha\beta}=\alpha'^2 k^2 \big(-\tfrac{k^2}{2}V^{\alpha\beta}-\tfrac{1}{2}k.V.k \eta^{\alpha\beta}+k^{\alpha} k.V^{\beta}\big)\label{S2}\,.
\end{equation}
Using this source and \eqref{V0} we get
\begin{eqnarray}
&&{\cal A}=\frac{-t^2}{t}(k_1 \cdotp V \cdotp {\epsilon_2} \cdotp {\epsilon_1} \cdotp k_2 
-k_1 \cdotp V \cdotp {\epsilon_1} \cdotp {\epsilon_2} \cdotp k_1    
+ \tfrac{1}{2}  k_1\cdotp {\epsilon_2}\cdotp k_1 Tr( \epsilon_1 \cdotp V) - k_1\cdotp {\epsilon_2}\cdotp V \cdotp      {\epsilon_1}\cdotp k_2  \nonumber \\ 
&& + \tfrac{1}{2} k_2 \cdotp {\epsilon_1}\cdot k_2 Tr(\epsilon_2\cdotp V) 
-  \tfrac{1}{2} s\, Tr(\epsilon_1 \cdotp {\epsilon_2})  -  \tfrac{1}{2} t\, Tr(\epsilon_1 \cdotp V \cdotp {\epsilon_2}))\,, \label{ATT12}
\end{eqnarray}
which again cancels  the second order expansion (contact terms) of $\square R_{\alpha\beta} \tilde{G}^{\alpha\beta}$. 
  
 So eventually  computation of effective field theory amplitude leads to the string theory t-channel $ {\cal A}_{t}(\alpha'^2) $ plus  some contact terms $\mathcal{C}_t$ that must be considered  in the case of contact interactions
\begin{eqnarray}\label{CTE}
&&\mathcal{C}_t=\tfrac{1}{32}\big( s t Tr(\epsilon_1. \epsilon_2) - 2 t k_2.{\epsilon_1}. {\epsilon_2}.V.k_1  -4 (k_2.{\epsilon_1}.k_2)(k_1.{\epsilon_2}.V.k_1)-  t^2 Tr(\epsilon_2. V.\epsilon_1)\nonumber \\ 
&& + 2 t k_1.\epsilon_2.\epsilon_1.V.k_1 -4 (k_1.{\epsilon_2}.k_1)(k_2.{\epsilon_1}.V.k_2) -  2k_2.\e_1.V.\e_2.k_1\big)\,.
\end{eqnarray} 

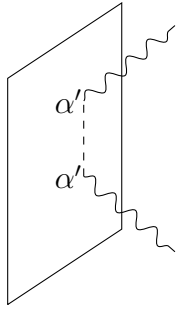
\begin{figure}
 	\centering
 	\begin{tikzpicture}
 	\draw (1.5,0) -- (3,1) -- (3,4) -- (1.5,3) -- (1.5,0);
 	\draw [decoration={snake},decorate] 
 	(2.5,2.7) -- (3.7,3.7);
 	\draw [decoration={snake},decorate] 
 	(2.5,1.7) -- (3.7,.7);
 	\draw [dashed]  (2.5,1.7) -- (2.5,2.7);
 	\node at (2.3,1.7) {$\alpha'$};
 	\node at (2.3,2.7) {$\alpha'$};
 	\end{tikzpicture}
 	
 	\caption{{\small s-channel diagram with scalar field that propagates on the brane.}}\label{sch}
 \end{figure}

The field theory diagram for amplitude in $s$-channel is shown in (Fig.\ref{sch}). For amplitude to be at $O(\alpha'^2)$, each of vertices between open and closed string fields must be at $O(\alpha')$. Note that it can not be possible to have one vertex at $O(\alpha'^2)$ and the other at $O((\alpha')^0)$ because
 in  \cite{Ardalan:2002qt} it has been shown that the closed-open interaction
  in Bosonic string theory  are maximally at order $\alpha'$. 
  By considering this point the s-channel amplitude  becomes 
  \begin{equation}
  \mathcal{A}_s(\e_1\e_2)=(\mathcal{V}^{\alpha'}_{\e_1\lambda})^{i}  (\mathcal{P}_{\lambda\lambda})_{ij} (\mathcal{V}^{\alpha'}_{\lambda\e_2})^{j} + (\e_2\leftrightarrow\e_1)\,,\label{ASf}
  \end{equation}
 where $  (\mathcal{P}_{\lambda\lambda})_{ij} $ is the propagator for open scalar fields $\lambda^i$  and $ (\mathcal{V}^{\alpha'}_{\e_1\lambda})^{i} $ is the $O(\alpha')$ open-closed vertex on the $D_p$-brane which must be computed from the action (\ref{dbi1}). 
 Computing this amplitude also gives us the string theory s-channel in Eq.(\ref{ASo2}) plus some contact terms $ {\cal C}_{s}$
\begin{eqnarray}\label{CSE}
 &&\mathcal{C}_s=-\tfrac{1}{8}\big(s (2 s + t) Tr(\e_1.V) Tr(\e_2.V)+ (2 s + t) Tr(\e_1.V) k_1.V.\e_2.V.k_1 \nonumber \\ 
 &&\qquad+ (2 s + t) Tr(\e_2.V) k_2.V.\e_1.V.k_2  + 2 (k_1.V.\e_1.V.k_1)(k_1.V.\e_2.V.k_1) \big)\,.
 \end{eqnarray}
 
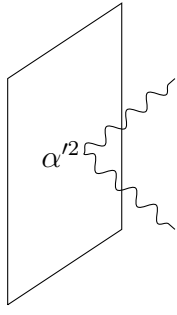
\begin{figure}
    \centering
\begin{tikzpicture}
\draw (1.5,0) -- (3,1) -- (3,4) -- (1.5,3) -- (1.5,0);
\draw [decoration={snake},decorate] 
(2.5,2) -- (3.7,3);
\draw [decoration={snake},decorate] 
(2.5,2) -- (3.7,1);
\node at (2.2,2) {$\alpha'^2$};
\end{tikzpicture}
\caption{Contact interaction of bulk and D-brane at $O(\alpha'^2)$.}\label{cnt}
\end{figure}

Until now our calculations have not required adding extra $\alpha'^2$ terms to the action and we have just shown the consistency of previously known actions in the bulk and $D_p$-brane.  
Here we prove that for consistency of the field theory contact terms with the string theory amplitude we need to add  $\alpha'^2R^2$ terms to the $D_p$-brane action. Some of these contact terms already come from (Fig.\ref{talpha2}) and (Fig.\ref{sch}) or equivalently from equations (\ref{CTE}) and (\ref{CSE}). But in general we need to add extra terms to the action 
for compensation of the diagram in (Fig.\ref{cnt}).
 These terms must be quadratic in graviton field and have four number of derivatives. To find the correct coupling we consider the most general form of such corrections
\begin{align}
&\qquad\qquad\,\,\,\,\,\,\mathcal{L}_1=\r_1 R_{\alpha \delta}{}^{\lambda \mu} R_{\beta \kappa \lambda \mu} \tilde{G}^{\alpha \beta} \tilde{G}^{\delta \kappa}  + \r_2 R_{\alpha}{}^{\lambda}{}_{\delta}{}^{\mu} R_{\beta \lambda \kappa \mu} \tilde{G}^{\alpha \beta} \tilde{G}^{\delta \kappa}+ \r_3 R_{\alpha}{}^{\lambda}{}_{\beta}{}^{\mu} R_{\delta \lambda \kappa \mu} \tilde{G}^{\alpha \beta} \tilde{G}^{\delta \kappa}\nonumber  \\ 
&  + \r_4 R_{\alpha \delta \lambda}{}^{\nu} R_{\beta \kappa \mu \nu} \tilde{G}^{\alpha \beta} \tilde{G}^{\delta \kappa} \tilde{G}^{\lambda \mu}+ \r_5 R_{\alpha \delta \beta}{}^{\nu} R_{\kappa \lambda \mu \nu} \tilde{G}^{\alpha \beta} \tilde{G}^{\delta \kappa} \tilde{G}^{\lambda \mu}+ \r_6R_{\alpha \delta \lambda \nu} R_{\beta \kappa \mu \rho} \tilde{G}^{\alpha \beta} \tilde{G}^{\delta \kappa} \tilde{G}^{\lambda \mu} \tilde{G}^{\nu \rho} \nonumber  \\ 
& + \r_7 R_{\alpha \delta \beta \lambda} R_{\kappa \nu \mu \rho} \tilde{G}^{\alpha \beta} \tilde{G}^{\delta \kappa} \tilde{G}^{\lambda \mu} \tilde{G}^{\nu \rho} 
+ \r_8 R_{\alpha \delta \beta \kappa} R_{\lambda \nu \mu \rho} \tilde{G}^{\alpha \beta} \tilde{G}^{\delta \kappa} \tilde{G}^{\lambda \mu} \tilde{G}^{\nu \rho} \,. \label{BR2}
\end{align}
Here we have not considered terms containing the Ricci tensor because they can be produced by field redefinition by applying on the previous terms from lower orders and so are ambiguous.  
Now we must expand Eq.(\ref{BR2}) to second order of the metric perturbations and then go to the momentum space, we call the resulting terms by ${\cal C}_A$. The final step is  to impose  the following equality that must be hold between string theory and field theory contact terms
\begin{equation}
 {\cal A}_{c}(\alpha'^2)={\cal C}_A+{\cal C}_{s}+{\cal C}_{t}\,.
\end{equation}
This leads to the following values for the coefficients in (\ref{BR2})
\begin{align}
&\r_1= \tfrac{1}{48} (6 -  \pi^2)\,,\quad \r_2= \tfrac{1}{4}\,,\quad \r_3 = \tfrac{1}{24} \pi^2\,,\quad \r_4 = \tfrac{1}{24} (-12 + \pi^2)\,,\nonumber\\
&\qquad \r_5 = \tfrac{1}{12} \pi^2\,\quad \r_6 = \tfrac{1}{4} (1 -  \r_7)\,,\quad \r_8= \tfrac{1}{8} (-1 - 2 \r_7)\,.
\end{align}
Inserting these values into the Eq.(\ref{BR2}) we find the following effective gravitational action for Bosonic $D_p$-brane at order $\alpha'^2$
\begin{equation}\label{sbos2}
S_{Bos}^{(2)}=\frac{\pi^2\alpha'^2T_p}{48}\int d^{p+1}x\,e^{-\phi}\sqrt{-\tilde{G}} \mathcal{L}_{Sup}+\alpha'^2T_p\int d^{p+1}xe^{-\phi}\sqrt{-\tilde{G}}\, \mathcal{L}_B
\end{equation}
where 
\bea
{\cal L}_{Sup}\!\!\!\!&=&\!\!\!\!\tilde{G}^{\alpha \beta} \tilde{G}^{\nu \mu}\big(2 
 R_{\alpha}{}^{\rho}{}_{\beta}{}^{\kappa} R_ {\nu \rho \mu \kappa}- R_{\alpha \nu}{}^{\rho \kappa} R_ {\beta \mu \rho 
 	\kappa}  \big)\nn \\
	\!\!\!\!&-&\!\!\!\! 2 \tilde{G}^{\alpha \beta} \tilde{G}^{\nu \mu} \tilde{G}^{\rho \kappa} \big(2 R_{\alpha \nu \beta}{}^{\lambda} R_ {\rho\mu \kappa 
 	\lambda}- 
 R_{\alpha \nu \rho}{}^{\lambda} R_ {\beta \mu \kappa \lambda}  \big)\label{R2GG}\,,
\eea
is the Superstring Lagrangian and $ \mathcal{L}_B $ are the additional following terms without $\pi^2$ coefficients
\begin{align}\label{LB}
&\mathcal{L}_B= \tfrac{1}{8} R_{\alpha \delta}{}^{\lambda \mu} R_{\beta \kappa \lambda \mu} \tilde{G}^{\alpha \beta} \tilde{G}^{\delta \kappa}  + \tfrac{1}{4} R_{\alpha}{}^{\lambda}{}_{\delta}{}^{\mu} R_{\beta \lambda \kappa \mu} \tilde{G}^{\alpha \beta} \tilde{G}^{\delta \kappa}  -  \tfrac{1}{2} R_{\alpha \delta \lambda}{}^{\nu} R_{\beta \kappa \mu \nu} \tilde{G}^{\alpha \beta} \tilde{G}^{\delta \kappa} \tilde{G}^{\lambda \mu}\nonumber \\ 
& + \tfrac{1}{4} R_{\alpha \delta \lambda \nu} R_{\beta \kappa \mu \rho} \tilde{G}^{\alpha \beta} \tilde{G}^{\delta \kappa} \tilde{G}^{\lambda \mu} \tilde{G}^{\nu \rho}  -  \tfrac{1}{8} R_{\alpha \delta \beta \kappa} R_{\lambda \nu \mu \rho} \tilde{G}^{\alpha \beta} \tilde{G}^{\delta \kappa} \tilde{G}^{\lambda \mu} \tilde{G}^{\nu \rho}\,.
\end{align}
Note that there is also the following structure which its coefficient does not fixed
\begin{eqnarray}
&&- 
\frac{\r_7}{4}(R_{\alpha \delta \lambda \nu} R_{\beta \kappa \mu \rho} - 4 R_{\alpha \delta \beta \lambda} R_{\kappa \nu \mu \rho}  + R_{\alpha \delta \beta \kappa} R_{\lambda \nu \mu \rho}) \tilde{G}^{\alpha \beta} \tilde{G}^{\delta \kappa} \tilde{G}^{\lambda \mu} \tilde{G}^{\nu \rho}\,.
\end{eqnarray}
This coefficient remains ambiguous but  at second order of field expansion it is equal to the Gauss-Bonnet terms constructed from $D_p$-brane metric and is a total derivative. Using similar arguments as in \cite{Bachas:1999um} one may set  $ \r_7 $ to zero. 
 
The $\alpha^{\prime}$ corrections  are absent in Superstring $D_p$-brane action. In \cite{Ghodsi:2016qey} it has been shown that the $\alpha'^2R^2$ terms in Eq.(\ref{R2GG}) can be written as those in \cite{Bachas:1999um}  for totally geodesic embedding i.e. 
 \bea
S_{Sup}^{(2)}=\frac{\pi^2\alpha'^2T_p}{48}\!\int \! d^{p+1}x\,e^{-\phi}\sqrt{-\tilde{G}}\big(R_{abcd}R^{abcd}-2\hat{R}_{ab}\hat{R}^{ab}-R_{abij}R^{abij}+2\hat{R}_{ij}\hat{R}^{ij}\big)\,, \label{SupD2} 
 \eea
where  $\hat{R}_{ab}=\tilde{G}^{cd}R_{cadb}$ and $\hat{R}_{ij}=\tilde{G}^{cd}R_{cidj}$.
 
Now let's try to find those terms that are coming from the field redefinition.  By using the field redefinition in Eq.\eqref{Frdef1} and by a pullback we can find the field redefinition on $D_p$-brane
\be\label{red2}
\delta\tilde{G}_{ab}=\alpha'( R_{ab}-\frac{1}{2D-4} \tilde{G}_{ab}R)+\a'^2(-\frac34 \partial_{a}X^\mu \partial_{b}X^\nu \square R_{\mu\nu}+\frac{3(D-3)}{8(D-2)^2} \tilde{G}_{ab} \square R)\,.  
\ee
 Applying this on \eqref{dbi} plus \eqref{dbi1} we can get the following terms at $ \alpha'^2 $-order 
  \begin{eqnarray}\label{amb2}
  &&\delta \mathcal{L}=\tilde{R}^{a b} R_{a b} -  \tfrac{1}{2} \tilde{R} R_{\alpha\beta}\tilde{G}^{\alpha\beta} + \frac{(p-1) \tilde{R} R}{4 (D-2)} +\frac{(p^2-1) R^2}{64 (D-2)^2}-  \frac{(p-1) R_{\alpha \beta} R \tilde{G}^{\alpha \beta}}{16 (D-2)} \nonumber\\
  && -\frac{R_{\alpha \beta} R \tilde{G}^{\alpha \beta}}{4 (D-2)}-  \frac{(p-3) R R_{\alpha \mu \beta \nu} \tilde{G}^{\alpha \beta} \tilde{G}^{\mu \nu}}{8 (D-2)}-  \tfrac{1}{8} R_{\alpha \gamma} R_{\beta \delta} \tilde{G}^{\alpha \beta} \tilde{G}^{\gamma \delta} + \tfrac{1}{16} R_{\alpha \beta} R_{\gamma \delta} \tilde{G}^{\alpha \beta} \tilde{G}^{\gamma \delta}\nonumber\\
  &&  -  R_{\gamma \delta} R_{\mu \alpha \nu \beta} \tilde{G}^{\alpha \beta} \tilde{G}^{\gamma \mu} \tilde{G}^{\delta \nu} 
	+ \tfrac{1}{4} R_{\gamma \delta} R_{\mu \alpha \nu \beta} \tilde{G}^{\alpha \beta} \tilde{G}^{\gamma \delta} \tilde{G}^{\mu \nu} + \tfrac{1}{2} g^{\beta \gamma} R_{\delta \gamma} R_{\beta \nu \mu \alpha} \tilde{G}^{\delta \mu} \tilde{G}^{\nu \alpha}\nonumber \\ 
  && + \frac{p \tilde{G}^{\alpha \beta} \nabla_{\beta}\nabla_{\alpha}R}{4 (D-2)}  + \tfrac{1}{2} \tilde{G}^{\alpha \beta} \tilde{G}^{\mu \nu} \nabla_{\nu}\nabla_{\beta}R_{\alpha \mu} -  \tfrac{1}{2} \tilde{G}^{\alpha \beta} \tilde{G}^{\mu \nu} \nabla_{\nu}\nabla_{\mu}R_{\alpha \beta}  \nn\\
	&&  - \tfrac{3}{8}  \square R_{\mu\nu}\tilde{G}^{\mu \nu} + \frac{3 (D-3) }{16 (D-2)^2}(p+1) \square R\,.
  \end{eqnarray}
These terms have not any effect on contact terms at $\a'^2$-order. All terms in the fourth line in above equation become total derivative at linear perturbation and the last two terms of the fifth line give rise to both t-channel and contact term where cancel each other. The remaining terms in the first three lines and those in the fourth line have at least one bulk Ricci term and consequently have not any effect for on-shell scattering of two gravitons.

\section{Summary and Discussion}\label{dis} 

 In this paper in introduction we reviewed the effective gravitational action on the bulk space-time. We observed that to have a ghost-free theory we need to fix the coefficients of the field redefinition (\ref{Frdef}). This gives rise to the Gauss-Bonnet action at $\a'$-order (\ref{GB}). Applying the field redefinition of (\ref{Frdef}) to the effective action produced by computing the string S-matrix amplitude (\ref{easmatrix1}) will give new contribution to $\a'^2$-order and it ruins the ghost-free condition at this order. To get ride of ghosts again we need to add new terms to the field redefinition at $\a'^2$-order i.e. (\ref{Frdef1}). This last field redefinition gives the final form of the effective action in the bulk at $\a'^2$-order, Eq.(\ref{bulk2}). An interesting observation at the level of Feynman graphs happens and one can show that the exchange diagram cancel by contact diagram (see (Fig.\ref{bulk2g})) for those terms that created by field redefinition.

In computing the effective gravitational actions on $D_p$-branes one expects that the above mentioned field redefinition is important at each order of $\a'$ expansion. In section 2 we show this by starting from the known result found in \cite{Corley:2001hg}, computed from S-matrix amplitude of scattering of two gravitons from $D_p$-branes (\ref{dbi1}). To impose the effect of field redefinition it is enough to induce the field redefinition of the bulk  (\ref{Frdef}) into the $D_p$-branes space-time i.e. Eq.(\ref{Frdefbrane}). Applying this to the DBI action at zero order (\ref{dbi}), will produce new extra terms to those which have been found in \cite{Corley:2001hg}. These new terms are presented in Eq.(\ref{EXlag1}). Again one can see the cancellation between different Feynman diagrams for ambiguous terms here. For $D_p$-branes this cancellation happens between t-channel diagram and contact term diagram (see (Fig.\ref{DbraneAmbg})).
A similar cancellation between exchange and contact diagrams for open string case in Superstring theory has been observed in \cite{Fotopoulos:2001pt}.

We perform the same computation in section 3 but we notice that unlike the previous order we have not the effective action at $\a'^2$-order. Therefore we need to compute the expansion of S-matrix amplitude of two gravitons from $D_p$-branes at this order which includes t-channel, s-channel and contact terms (\ref{ATo2}) - (\ref{AC1}). This calculation also support
the conjecture of uniform transcendentality in string theories in \cite{Huang:2016tag}. The details of field theory computations are presented in subsection 3.2. The final results without considering the field redefinition are given in Eqs. (\ref{sbos2}) - (\ref{LB}). But there are ambiguous terms again which do not produce by the S-matrix calculations. These terms are given in Eq.(\ref{amb2}) due to the field redefinition of (\ref{red2}). These new terms divide into two parts, those which have no effect for on-shell scattering of two gravitons and those which have  cancellation between t-channel diagram and contact term diagram.

As we mentioned before the existence of field redefinitions are crucial for gravitational theory in the bulk to be ghost-free. If one uses the effective gravitational action on $D_p$-branes then the ghost-free condition needs to be considered. As an example  in \cite{Hassan:2010ys}  the D-brane induced gravity is used for addressing the cosmological constant problem  and they have studied the ghost instability in this regard. Our computations indicate that the new terms we found here must be taken into account for such applications \cite{WIP}. 
\section*{Acknowledgment}\addcontentsline{toc}{section}{Acknowledgment}

Almost all of the calculations in this study have been carried out by the Mathematica package xAct\cite{Nutma:2013zea}. A. G. would like to thank M. R. Garousi for valuable discussions. The work of A.G. is supported by Ferdowsi University of Mashhad under the grant 2/41780.

\appendix
\section{}
The kinematic factors for two gravitons scattering from $D_p$-branes are given by
\bea
d_1 & = &  k_2 \cdotp \e_1 \cdotp \e_2 \cdotp k_1 
- k_2 \cdotp \e_1 \cdotp D \cdotp \e_2 \cdotp k_1 - 2 k_2 \cdotp \e_1 \cdotp \e_2 \cdotp D \cdotp k_2 
 + (\mbox{1} \leftrightarrow
\mbox{2}), \nonumber\\
d_2 & = & \mbox{Tr}(\e_1 \cdotp D) (k_1 \cdotp \e_2 \cdotp k_1) +(\mbox{1} \leftrightarrow
\mbox{2}), \nonumber\\
d_3 & = & k_1 \cdotp D \cdotp \e_1 D \cdotp \e_2 \cdotp D \cdotp k_2  +  k_1 \cdotp D \cdotp \e_1 \cdotp \e_2 \cdotp D \cdotp k_2  + \mbox{Tr}(\e_1 \cdotp D)
(k_1 \cdotp D \cdotp \e_2 \cdotp D \cdotp k_1)  +(\mbox{1} \leftrightarrow
\mbox{2}), \nonumber\\
d_4 & = & \mbox{Tr}(\e_1 \cdotp D \cdotp \e_2 \cdotp D) + k_1 \cdotp D \e_2 \cdotp D \cdotp \e_1 \cdotp D \cdotp k_2
+ k_2 \cdotp D \cdotp \e_1 \cdotp D \cdotp \e_2 \cdotp D \cdotp k_1, \nonumber\\
d_5 & = & \mbox{Tr}(\e_1 \cdotp \e_{2}) - k_1 \cdotp \e_2 \cdotp \e_1 \cdotp k_2
- k_2 \cdotp \e_1 \cdotp \e_2 \cdotp k_1, \nonumber\\
d_6 & = & \frac{1}{2} \mbox{Tr}(\e_1 \cdotp D) \mbox{Tr}(\e_2 \cdotp D)
+ \mbox{Tr}(\e_1 \cdotp D) (k_2 \cdotp D \cdotp \e_2 \cdotp D \cdotp k_2) + (k_2 \cdotp \e_1 \cdotp k_2) 
(k_1 \cdotp D \e_2 \cdotp D \cdotp k_1) \nonumber \\
& + & \frac{1}{2} (k_2 \cdotp \e_1 \cdotp D \cdotp k_2)(k_1 \cdotp \e_2 \cdotp D \cdotp k_1)
+ (k_2 \cdotp \e_1 \cdotp D \cdotp k_2)(k_1 \cdotp D \cdotp \e_2 \cdotp k_1) \nonumber \\
& + & \frac{1}{2} (k_2 \cdotp D \cdotp \e_1 \cdotp k_2)
(k_1 \cdotp D \cdotp \e_2 \cdotp k_1)  +  (\mbox{1} \leftrightarrow
\mbox{2}), \nonumber \\
d_7 & = & (k_1 \cdotp \e_2 \cdotp k_1) (k_2 \cdotp \e_1 \cdotp k_2), \nonumber \\
d_8 & = & (k_2 \cdotp \e_1 \cdotp k_2) (k_1 \cdotp \e_2 \cdotp D \cdotp k_1 + k_1 \cdotp D \cdotp \e_2 \cdotp k_1)
+ (\mbox{1} \leftrightarrow \mbox{2}), \nonumber \\
d_9 & = & (k_2 \cdotp D \cdotp \e_1 \cdotp D \cdotp k_2) (k_1 \cdotp \e_2 \cdotp D \cdotp k_1 + k_1 \cdotp D \cdotp \e_2 \cdotp k_1)
+ (\mbox{1} \leftrightarrow \mbox{2}), \nonumber \\
d_{10} & = & (k_1 \cdotp D \cdotp \e_2 \cdotp D \cdotp k_1) (k_2 \cdotp D \cdotp \e_1 \cdotp D \cdotp k_2). 
\eea

\providecommand{\href}[2]{#2}\begingroup\raggedright

\endgroup
\end{document}